# Geometric origin of Eliott relation.

Yuri A. Serebrennikov

Qubit Technology Center

2152 Merokee Dr., Merrick, NY 11566

ys455@columbia.edu

It has been shown that the Elliott relation results from the random acquisition of geometric phases and represents a special case of a more general situation - relaxation of the pseudo spin-1/2 induced by stochastic gauge fields.

03.65.Ge , 71.70.Ej, 75.10.Jm

50 years ago Elliott[1] established that in the absence of external magnetic field the rate of spin relaxation in nonmagnetic semiconductors and metals, $1/T_S$, is proportional to the rate of the relaxation of the crystal momentum, $1/\tau_p$:

$$T_S^{-1} = a \, \| \delta\tilde{g}^2 \| \, \tau_p^{-1}, \qquad (1)$$

where $\delta\tilde{g} = \tilde{g} - g_e \hat{1}$ is the deviation of the "g-tensor" from the free-electron value, $a$ is a constant in the range 1 - 10. The relation (1) was obtained in the tight-binding approximation by introduction the spin-orbit interaction $H_{SO} = \lambda \vec{L}\cdot\vec{S}$, where $\vec{S}$ and $\vec{L}$ denote the electron spin and orbital momentum, into the crystal Hamiltonian. The presence of SOC leads to a mixing of spin-up and spin-down Bloch functions $|\vec{\sigma}, \vec{k}\rangle := |\vec{\sigma}, k(\vec{\Omega})\rangle$ in the different bands with the same crystal momentum $\vec{k} := k(\vec{\Omega})$,

where $\vec{\sigma}$ is the vector of Pauli matrices and $\vec{\Omega}$ represents the set of parameters necessary to uniquely identify the orientation of $\vec{k}$. Due to this mixing the *adiabatic* ($\Delta E \tau_p \gg 1$) scattering of a wave vector $\vec{k} \to \vec{k}'$ by fluctuations of the anisotropic part of electron-lattice Coulomb interaction results in a nonzero spin-flip probability $P(\vec{\sigma}, \vec{k} \to \vec{\sigma}', \vec{k}') \sim \sin^2 \vartheta_{\vec{k},\vec{k}'}$, where $\vartheta_{\vec{k},\vec{k}'}$ is the angle between $\vec{k}$ and $\vec{k}'$. The formula (1) is valid to the first order of $\lambda/\Delta E \ll 1$, where $\Delta E$ is the energy separation from the considered band state to the nearest one with the same $\vec{k}$, and $\lambda$ characterize the amplitude of the matrix element of $H_{SO}$ between these states. The Elliott relation passed the experimental tests for many alkali and noble metals, and bulk semiconductors with relatively small interband gap (to satisfy adiabatic conditions $\Delta E$ should be larger than $\tau_p^{-1}$) and large SOC.

Although adiabatic motions do not change the eigenvalues of the Hamiltonian, they affect the eigenvectors, thereby giving rise to phase shifts and to transitions between quantum states that are *geometric* in nature[2]. The nontrivial gauge potentials, either Abelian or non-Abelian, which appear in systems that undergo slow loops in the parameter space lead to a geometric phase of a state vector[2]. It has been shown[3,4,5,6,7] that if the adiabatic rotation of an external electric field is continuous and coherent, then the transition phase shift that is equal to the difference between the phase shifts acquired by the components of the KD, will increase linearly in time, which is equivalent to spin-precession or zero-field splitting[8]. These transitions between the doubly degenerate Kramers states depend only on the *trajectory* of the electric field in the angular space. Thereby an intriguing connection between the gauge fields and the specific form of spin-electric coupling: spin-rotation interaction was revealed. Note that revolving electric field violates the T-invariance of a system and hence there is no contradiction to Kramers theorem. If adiabatic rotations are incoherent, the resulting transition phase shift will be random and could lead to dephasing, which is equivalent to relaxation of the pseudo spin-1/2 (due to SOC spin is not a good quantum number) that represents the KD in zero-magnetic field[9][10]. Due to enormous generality and non-model character of the geometric results, many old problems appear in a new light. Here, we will show that Elliot relation arises from the random acquisition of geometric phases and represents a special case of a more general situation - relaxation of the pseudo spin-1/2 induced by stochastic gauge fields.

In the Elliott mechanism of spin relaxation the loss of coherence occurs only in the short time intervals during collisions. It is convenient to describe the particular scattering event in the moving (M) frame of reference that follows the adiabatic rotation of $\vec{k}$. In the absence of external magnetic fields, the quantization axis of the system coincides with the Z-axis of the M-frame that may be chosen along the direction of $\vec{k}$. At any instantaneous orientation of $\vec{k}(t)$, in zero magnetic fields (in crystals with inversion symmetry) the *s*-like electron Bloch function is doubly degenerate. The gauge transformation into the rotating M-basis leads to the following Schrödinger-type equation for the evolution of the Bloch KD adiabatically isolated from the rest of the band structure (for details see Refs.[3, 7]):

$$i\dot{\Psi}^{(M)}{}_{KD}(t) = H^{(M)}{}_{eff}(t)\Psi^{(M)}{}_{KD}(t), \qquad (2)$$

$$H^{(M)}{}_{eff}(t) := -1/2\,\vec{\omega}(t)\,\vec{\gamma}^{(M)}\vec{\sigma}^{(M)} = -i\,A^{(M)}{}_{WZ}(t). \qquad (3)$$

To simplify notations, $\hbar$ has been set equal to unity, and we introduce

$\Psi^{(M)}{}_{KD}(t) := |\vec{\sigma}^{(M)}, k[\vec{\Omega}(t)]; c>$, where $c$ is a conduction band index. The $A^{(M)}{}_{WZ}$ is the Wilczek-Zee non-Abelian gauge potential[2], and $\vec{\omega}(t)$ is an instantaneous angular velocity of the M-frame relative to the *space-fixed* lab (L) frame at time $t$. The "tensor" $\vec{\tilde{\gamma}}^{(M)}$ is defined by the expression [10]

$$1/2\, \vec{\tilde{\gamma}}^{(M)}\, \vec{\sigma}^{(M)} := P^{(M)}{}_{KD}[R^{-1}(t)\vec{J}^{(L)}R(t)]P^{(M)}{}_{KD}, \qquad (4)$$

where $P^{(M)}{}_{KD}$ is the projector onto the complex two-dimensional (*2D*) Hilbert space spanned by the Kramers doublet. The Schrödinger-type equation (2) and the expression (3) depend on a choice of gauge that specifies the *reference* orientation, i.e. the orientation in which the M-frame coincides with some space-fixed frame. At the moment $t = 0$, the reference orientation may always be chosen such that $\vec{\tilde{\gamma}}^{(M)}$ is diagonal and that the main axis $Z$ of this "tensor" represents the quantization axis of the pseudo spin operator $\vec{S}^{(M)}{}_{eff} := \vec{\sigma}^{(M)}/2$. The rotation operator $R = \exp[-i\phi \hat{\vec{n}}\vec{J}]$ in Eq.(4) maps the space-fixed reference orientation into the actual orientation of the M-frame at time $t$, $\Psi^{(M)}(t) = R(t)\Psi^{(L)}(t)$, where $\Psi$ is the instantaneous adiabatic eigenvector of the total, *nontruncated* Hamiltonian of the crystal, $H$. Recall that in the presence of SOC, the Bloch functions are not factorizable into the orbital and spin parts, hence, the total electron angular momentum, $\vec{J} = \vec{L} + \vec{S}$, is included into the transformation $R(t)$. The unit vector $\vec{n}$ denotes the instantaneous axis of the $\vec{k} \to \vec{k}'$ rotation and the angle of this rotation is denoted as $\phi$. Clearly, this transformation is the gauge transformation and is responsible for the appearance of the gauge potential $A^{(M)}_{WZ}(t)$ in the Eq.(3).

Suppose that during a short time of a collision, $\delta t_c$, the plain of the $\vec{k} \to \vec{k}'$ rotation remains constant. In this situation, the gauge potential $A_{WZ}^{(M)}(t)$ lost its non-Abelian character and Eq.(2) is readily integrable:

$$\vec{u}_Z^{(M)}(t) = Tr\{\vec{\sigma}_Z^{(M)} \exp[(i\omega t \gamma_\perp / 2)\vec{\sigma}_X] \vec{\sigma}_Z^{(M)}\} = \cos(\omega t \gamma_\perp),$$

where the axis of rotation $\vec{n}$ was assigned to $X$, $\gamma_\perp := \tilde{\gamma}_{XX}^{(M)} = \tilde{\gamma}_{YY}^{(M)}$, and we introduce the polarization vector, $\vec{u}^{(M)}(t) := Tr[\rho^{(M)}{}_{KD}(t)\vec{\sigma}^{(M)}]$, $\rho^{(M)}{}_{KD}(t) := |\Psi^{(M)}{}_{KD}(t)\rangle\langle\Psi^{(M)}{}_{KD}(t)|$ is the corresponding density operator. To describe the evolution of the KD during a collision in the *local* reference frame we have to perform a reverse rotation of the basis compensating for the rotation of the M-frame, which yields the following result ($\delta\gamma_\perp := \gamma_\perp - 1$):

$$\vec{u}_Z^{(L)}(t) = Tr\{\vec{\sigma}_Z^{(L)} \exp[(i\omega t \delta\gamma_\perp / 2)\vec{\sigma}_X] \vec{\sigma}_Z^{(L)}\} = \cos(\omega t \delta\gamma_\perp). \qquad (5)$$

We would like to emphasize that Eq.(5) is applicable only during the collision ($t \le \delta t_c, \omega t \le \pi$), in the local reference frame that reflects the geometry of the *particular* scattering event.

This simple form can be easily rationalized. The effective spin-Hamiltonian $H^{(M)}{}_{eff}$, Eq.(3), can be viewed as a generic Zeeman Hamiltonian of a spin-1/2 particle in an "effective" time-dependent magnetic field $\vec{\omega}(t)\tilde{\gamma}^{(M)}$ that appears during the collision in the frame that follows the adiabatic rotation of $\vec{k}$. In the local reference frame the differential *action* of $H^{(L)}{}_{eff}$ is proportional to the angle of rotation, $|\vec{\omega}(t)(\tilde{\gamma}^{(M)} - \hat{1})| dt$, i.e., to the *distance* in the angular space, which reveals the geometric character of the phenomenon. Accordingly, as long as the reorientation of a crystal momentum represents an adiabatic perturbation to the system, the evolution of the spinor $\Psi^{(L)}{}_{KD}$ depends only on the path traveled by $\vec{k}$ in the angular space and is independent of the way the system moves along that path. In other words, the geometric Berry-phase shift acquired by the spin-up and spin-down components of the KD during a collision depends only on the angular distance, $\vartheta_{\vec{k},\vec{k}'} = \omega \delta t_c$. The polarization vector follows the reorientation of the lattice momentum, but is generally falling somewhat behind ($\delta\gamma_\perp \vartheta_{\vec{k},\vec{k}'}$).

Now we are ready to address the effect of random collisions that result in a stochastic acquisition of Berry's phase. First, we consider 2D crystals, where the axis $X$ of the local reference frame ($\vec{n}$) is constant and can be chosen to be at right angle to the lateral plane. Suppose that an average angle of the in-plain $\vec{k} \to \vec{k}'$ rotation is small ($\vartheta_{\vec{k},\vec{k}'} \ll 1$), then the stochastic scattering process of a wave vector can be safely modeled by the one-dimensional diffusion in the angular space. In this case, the gauge potential in Eq.(3) is "Abelianized" and Eq.(5) can be easily averaged, <...>, over the stochastic ensemble with the probability density function $P(\vartheta_{\vec{k},\vec{k}'},t) = (4\pi Dt)^{-1/2} \exp(-\vartheta^2_{\vec{k},\vec{k}'}/4D_1 t)$. Integration over all possible angles $\vartheta_{\vec{k},\vec{k}'}$ gives $<\vec{u}_Z^{(L)}(t)> = \exp(-\delta\gamma_\perp^2 D_1 t)$, i.e., an exponential decay of spin coherence with the rate

$$1/T_S(2D) = \delta\gamma_\perp^2 D_1, \qquad (6)$$

proportional to the one-dimensional diffusion coefficient $D_1$. The 3D case is generally more complex since $\vec{n}$ can change its direction in time, so the elementary rotations in the local and the mesoscopic reference frames may not commute. We will return to this point below.

Note that all relevant information about the crystal Hamiltonian, which comprises the actual physical problem, is now represented by the "$\tilde{\gamma}$-tensor". The original, nontruncated $H$ serves only to determine the gauge group and the principal values of $\tilde{\gamma}^{(M)}$. The explicit form of $P^{(M)}{}_{KD}$ and, thus, $\tilde{\gamma}^{(M)}$ depends on the problem at hand. Examples of "$\tilde{\gamma}$-tensor" calculations can be found in Refs. [10]. It has been shown[7], that generally in the case of weak SOC to the first order in $\lambda/\Delta E$,

$$\tilde{\gamma}^{(M)} - \hat{1} = \Delta\vec{g}^{(M)}.$$

In this situation, it is natural to assume that $|\vec{\omega}(t)(\tilde{\gamma}^{(M)} - \hat{1})|\delta t_c \ll 1$, even for $\vartheta_{\vec{k},\vec{k}'} \sim 1$. Consequently, the response of the pseudo spin is much slower than $\tau_p$ or, equivalently, the rate of transitions between $m = \pm 1/2$ levels of the Bloch KD, is much slower than the inverse correlation time of the fluctuating effective Hamiltonian, Eq.(3). From the geometric point of view, this means that on the average the state of a quantum system, represented by $<\vec{u}(t)>$, is independent of the particular position in the angular space. Thus, the problem reduces to the traditional calculations of the relaxation operator in the "fast motional" limit that allows us to carve the following result (see[11] for details):

$$1/T_S(3D) = 4/3\,\delta\gamma_\perp^2 <\omega^2> \tau_c := 4/3\Delta g_\perp^2 \tau_p^{-1}, \qquad (7)$$

where $\tau_c$ is the correlation time of $\vec{\omega}$.

Our results, Eqs.(6) and (7) have a non-model geometric origin and do not depend on the system being electronic or nuclear. The geometric dephasing in the context of zero-field NQR experiments was first considered by Jones and Pines[10], who derived a decoherence rate of $^{131}$Xe nuclei induced by thermal collisions with the walls of a toroidal container. They start from the dynamic evolution of the *nuclear* KD adiabatically isolated from the rest of a spin-multiplet (nuclear spin $I > \frac{1}{2}$), which can be described by Eqs.(2) and (3) with $\gamma_\parallel = 1$, $\gamma_\perp = I + 1/2$ (see Refs.[2] for details). It is easy to see that for $^{131}$Xe ($I = 3/2$) nuclear pseudo spin just follows the rotation of the M-frame, which in this case coincides with the main axes of the tensor of quadrupolar interaction: $H_{eff}^{(L)}(t)dt = -[\vec{\omega}_X(t)\vec{\sigma}_X^{(L)} + \vec{\omega}_Y(t)\vec{\sigma}_Y^{(L)}]dt/2$. The one-dimensional diffusion model utilized by Jones and Pines then yields $1/T_{I=3/2,|m|=1/2}(2D) = D_1$, which represents another special case of Eq.(6).

The way we derived Eq.(7) reveals the geometric origin of Elliott relation, Eq.(1). It is fundamentally connected to the admixture of "fast" electron spatial degrees of freedom to spin wave functions. The states that are coupled by SOC to form the Bloch KD have energy separations of electron interband excitation. Therefore, high frequencies will characterize the time dependent response of the system to motionally induced perturbation acting towards the change in the mixing coefficients of the zero order wave functions. As a result, during the "slow" scattering event electron pseudo spin will adiabatically follow the rotation of the crystal momentum. Generally, this effect can be described as a manifestation of a relevant gauge potential and can be represented in purely geometric terms as a consequence of the corresponding geometric connection[2]

---

[1] R. J. Elliott, Phys. Rev. **96**, 266 (1954).

[2] M. V. Berry, Proc. R. Soc. London Ser. A **392**, 45 (1984); F. Wilczek and A Zee, Phys. Rev. Lett. **52**, 2111 (1984); A. Zee, Phys. Rev. A **38**, 1 (1988); C. A. Mead, Phys. Rev. Lett. **59**, 161 (1987); J. Segert, J. Math. Phys. **28**, 2102 (1987).

[3] Yu. A. Serebrennikov, Phys. Rev. B. **70**, 064422 (2004).

[4] B. A. Bernevig and S.-C. Zhang, arXiv: quant-ph/0402165 (2004).

[5] A. V. Balatsky and B. L. Altshuler, Phys. Rev. Lett. **70**, 1678 (1993).